\documentclass[11pt,amssymb,preprint,aps]{revtex4}
\usepackage{mathrsfs}
\usepackage{amsfonts}
\usepackage{amsmath}
\usepackage{graphicx}
\newcommand{\bbe}{\begin{equation}}
\newcommand{\be}{\begin{equation}}
\newcommand{\ee}{\end{equation}}

\def \cM {{\cal M}}
\def \cN {{\cal N}}
\def \cT {{\cal T}}
\def \cL {{\cal L}}
\def \oM {\overline{\phantom A}\hspace{-3.6mm}M}
\def\sqr#1#2{{\vcenter{\vbox{\hrule height.#2pt\hbox{\vrule width.#2pt  
height#1pt \kern#1pt\vrule width.#2pt}\hrule height.#2pt}}}}
\begin{document}

\title{ 
The Cauchy problem on spacetimes that are not globally hyperbolic  
} 
 
\author{John L. Friedman} 
\affiliation{ 
Department of Physics, University of Wisconsin-Milwaukee, P.O. Box 413,  
Milwaukee, WI 53201, U.S.A.} 

\begin{abstract}
 
The initial value problem is well-defined on a class of spacetimes
broader than the globally hyperbolic geometries for which existence and
uniqueness theorems are traditionally proved.  Simple examples are the
time-nonorientable spacetimes whose orientable double cover is globally
hyperbolic. These spacetimes have generalized Cauchy surfaces on which
smooth initial data sets yield unique solutions.  A more difficult
problem is to characterize the class of spacetimes with closed timelike
curves that admit a well-posed initial value problem.  Examples of
spacetimes with closed timelike curves are given for which smooth
initial data at past null infinity has been recently shown to yield
solutions.  These solutions appear to be unique, and uniquesness has
been proved in particular cases.  Other examples, however, show that
confining closed timelike curves to compact regions is not sufficient
to guarantee uniqueness.  An approach to the characterization problem
is suggested by the behavior of congruences of null rays.  Interacting
fields have not yet been studied, but particle models suggest that
uniqueness (and possibly existence) is likely to be lost as the
strength of the interaction increases.

\end{abstract} 
\maketitle 
 
\section{Introduction} 
\label{intro}

Motivating the definition of a {\em globally hyperbolic} spacetime are two
facts:  On globally hyperbolic spacetimes, wave equations have a
well-defined initial value formulation; and the ordinary causal
structure of a globally hyperbolic spacetime mirrors the ordinary
causal structure observed in the universe.  The requirement that 
the initial value problem be well defined, however, picks out a 
broader class of spacetimes.  And the causal structure of the 
physical universe on the largest and smallest scales may 
not conform to that of a globally hyperbolic spacetime. 
  
In particular, a class of spacetimes that are not globally hyperbolic and 
nevertheless admit a well-defined initial value problem are Lorentzian 
4-geometries with a single spacelike boundary  -- Lorentzian 
{\it universes-from-nothing} (Friedman and Higuchi \cite{fh,f98}).  
These are the metrics that arise in a Lorentzian path-integral 
construction of the Hartle-Hawking wavefunction of the Universe.  
There is large class of such geometries, spacetimes that are 
compact on one side of a spacelike boundary.  A simple two-dimensional 
example, considered in more detail below, is a M\"obius strip with 
a flat Lorentz metric, for which the direction orthogonal to the median 
circle is timelike.  On each underlying manifold of such a spacetime, 
one can choose metrics that have no closed timelike curves 
and for which the boundary remains spacelike; time nonorientability is 
then their  only causal pathology.  With metrics so chosen, these
spacetimes provide the only examples of topology change in which one
has a smooth, nondegenerate Lorentzian metric without closed timelike
curves.  Instead, the spacetimes are time nonorientable.  The initial
value problem for these spacetimes is discussed in Sect. II.

The more difficult problem is to characterize the class of spacetimes
that have closed timelike curves (CTCs) and that nevertheless allow a
well-defined initial value problem for hyperbolic systems.  This brief
review outlines recent work that has been done in proving existence and 
uniqueness of solutions to the scalar wave equation on a class of 
spacetimes with closed timelike curves.  Sect. III introduces the 
subject with several two-dimensional examples that are easily 
analyzed and illustrate the obstacles to the existence of a 
generalized initial value problem.  These obstacles are less 
severe in four dimensions, and Sect. IV considers four dimensional 
examples in which existence and uniqueness theorems have been proved.
The 4-dimensional examples are quite restricted; Section V outlines the 
heuristic arguments that have been made for the existence of a well-defined
initial value problem in a much broader class of spacetimes.  
The section concludes with two conjectures that would partly characterize
such a class.  

Let us first generalize the definition of a Cauchy surface
to allow spacetimes that are not globally hyperbolic. Let $M,g$ be 
a smooth spacetime, a manifold $M$ together with a Lorentz-signature 
metric $g$. Recall that a set $S\subset M$ is {\em achronal} if no 
two points are timelike separated.\cite{hawkellis})\\ 
{\em Definition}. A {\em generalized Cauchy surface} $\Sigma$ is an 
achronal hypersurface of $M$ for which the initial value problem for 
the scalar wave equation is well-defined: \\
For any smooth data in $L_2(\Sigma)$ with finite energy, for a scalar
field $\Phi$, there is a unique solution $\Phi$ on $M$.  

One can then ask what the class is of spacetimes for which the initial
value problem is well-defined.  To see that this class is larger 
than the class of globally hyperbolic spacetimes, we begin with the 
time non-orientable spacetimes mentioned above. 

\section {Time-nonorientable spacetimes with a well-defined initial 
value problem} 

We begin with a two dimensional example and then describe a general
class of geometries on a countably infinite set of 4-manifolds.
Consider the M\"obius strip $M$ with a metric for which its median
circle is spacelike.  One can construct such a 2-dimensional spacetime
from a cylinder $\oM=  \mathbb R\times S^1$, with an obvious choice of
Minkowski metric that makes copies of $\mathbb R$ into timelike lines
orthogonal to copies of $S^1$.  In terms of the natural chart $t,\phi$,
on $\oM$, the metric is $-dt^2+a^2d\phi^2$, some length $a$.  To construct the
M\"obius strip, identify each point $(t,\phi)$, with its antipodal
point \be A(t,\phi) = (-t, \phi+\pi).\ee The M\"obius strip is the
quotient space $M = \oM/A$; because $A$ is an isometry, $M$ inherits a
flat Lorentzian metric for which its median circle $\Sigma$ (the image
of the circle $t=0$) is spacelike.

It is easy to see that the median circle $\Sigma$ is a generalized
Cauchy surface (as is any boosted image of it).  Initial data is a pair
$\Phi, \nabla\Phi$ (with $\nabla$ the 2-dimensional gradient) on
$\Sigma$.  Data on $\Sigma$  lifts to initial data on a Cauchy surface
$\overline\Sigma$ of the cylinder, and the data is antipodally
symmetric.  Because the the cylinder is globally hyperbolic, there is a
unique solution $\overline\Phi$ to the wave equation with this data,
and that solution is itself antipodally symmetric.  Thus there is a
field $\Phi$ on the M\"obius strip, whose lift to the cylinder
is $\overline \Phi$; and $\Phi$ is the unique solution to the wave equation
on $M$ with the specified initial data.

In this flat example, the M\"obius strip has closed timelike curves
(CTCs) {\em and} is time norientable.  If one chooses a deSitter 
metric on the cylinder instead of the flat metric, the antipodal 
map remains an isometry, and the M\"obius strip inherits a local deSitter 
metric, a metric for which it has no CTCs.  (CTCs arise from timelike
lines that emerge in opposite directions - forward and backwards 
with respect to a locally defined time direction;  in the deSitter 
geometry circles far from the median circle are large, and they 
expand fast enough that the timelike lines never meet.) 
In both the flat and the deSitter case (i.e., with and without CTCs),
because the orientable double-cover of $M,g$ is globally 
hyperbolic, the spacetime has a generalized Cauchy surface.  

More generally let $\overline\Sigma$ be any 3-manifold that admits a free  
involution, a diffeo $I$ that has no fixed points.  There are 
countably many spherical spaces and countably many hyperbolic 
spaces that admit such involutions. As in the above construction, 
one defines on a cylinder $\oM = \mathbb R\times \overline\Sigma$ an antipodal 
map $A=T\times I$,  where $T:\mathbb R\rightarrow \mathbb R$ is time reversal:  
\be 
A(t,p) = (-t, I(p)).
\ee
$A$ is again a free involution, and the manifold of the spacetime is again the 
quotient 
\be 
M = \oM/A.
\ee

We will choose a metric for which the 3-manifold $\Sigma =
\{0\}\times\overline\Sigma/A$ is a generalized Cauchy surface.  Any
metric $g$ on $M$, for which each of the surfaces corresponding to
$\{t\}\times\overline \Sigma$  are spacelike, will do.  The pullback of
$g$ to $\oM$ is a metric for which $M$ is foliated by the spacelike
hypersurfaces $\{t\}\times\overline \Sigma$ and is therefore globally
hyperbolic.  (For example, let  ${}^3g$ be any Riemannian metric on
$\Sigma$, ${}^3\overline g$ its pullback to $\oM$. The
metric $-dt^2 + {}^3\overline g$ on $\oM$ is antipodally symmetric
and induces an suitable metric on $M$.)

By our  construction of  $\overline g$, the antipodal map $A$ is an
isometry.  Again, the lift to $\overline \Sigma$ of data $\Phi,\nabla
\Phi$ on $\Sigma$ is an initial data set $\overline\Phi,
\nabla\overline\Phi$ that is antipodally symmetric (invariant under
$A$).  Because $\oM, \overline g$ is globally
hyperbolic, this data has a unique time evolution, $\overline \Phi$;
because both the data and the spacetime $\oM, \overline g$ are
antipodally symmetric, $\overline \Phi$ is antipodally symmetric.   The
field $\overline\Phi$  is therefore the lift to $\oM$ of a  solution
$\Phi$ to the scalar wave equation on $M$.  Finally, $\Phi$ is unique
because $\overline\Phi$ is unique. \\

Classically, spacetimes of the kind considered in this section (nonorientable 
spacetimes whose orientable double cover is globally hyperbolic), are 
locally indistinguishable from their covering spacetimes.  Treatments  
of classical spinor fields and of quantum field theory on such spacetimes 
are given in Refs. \cite{f95,gc,fh} \\

\section {Two-dimensional spacetimes with closed timelike curves}
\label{2d}

On spacetimes with CTCs, the initial value problem is more subtle.
Simple examples show that some spacetimes with CTCs have a generalized
Cauchy surface; but the Cauchy problem is not well-defined in generic
two-dimensional spacetimes, and other examples in this section
illustrate several essentially different ways in which CTCs can block
the existence of smooth solutions or allow more than one solution for
the same initial data on a spacelike surface. 

We again begin with two-dimensional spacetimes, built from Minkowski 
space.  Obstacles are most severe here, and we will see that 
some can be overcome in higher dimensions.
First a familiar case in which the one's naive expectation of the way 
CTCs prevent solutions is fullfilled.  Identify the edges of the strip 
of Minkowski space between two parallel, straight timelike lines, $t=0$ and 
$t=T$: 
\be
(t=0, x)\equiv(t=T, x).
\ee
Here the only candidates for an initial value surface are spacelike
lines $\Sigma$ extending to spatial infinity.  Minkowski space $\oM$
covers this spacetime, and data on any one of the spacelike lines of
$\oM$ that covers $\Sigma$ can be uniquely evolved in the covering
space, but the resulting solution corresponds to a solution on the
original spacetime only if it is suitably periodic. It must have the same 
value on each covering line $\overline\Sigma$; and almost no data
yields such a solution. This is essentially the grandfather paradox:
Locally one can construct a unique solution on $M$; but when extended, the 
locally evolved solution returns to $\Sigma$ with a value that is 
inconsistent with its initial data.

Flat cylinders with infinite extent in a timelike direction are obtained
by identifying the left and right edges of the strip of Minkowski
space between two parallel straight spacelike lines, $x=0$ and $x=d$, 
after a time translation by $\tau$:
\be
(t, x=0)\equiv(t+\tau, x=d).
\label{id}\ee
For $\tau> d$, lines joining identified points are CTC's.  
These spacetimes are everywhere dischronal:  A CTC passes 
through every point, and there is no candidate for an initial value 
surface -- no complete spacelike surface transverse to all timelike 
curves.  On spacetimes with no generalized Cauchy surface, one 
can ask a related question, more closely tied to our knowledge of 
the universe's causal structure:  whether the spacetime is {\it benign}
\cite{yurtsever94}.  A spacetime is benign if, at each point $x$, there 
is a finite spacelike surface $S$ containing $x$ for which arbitrary smooth 
data on $S$ can be extended to a solution on the spacetime.  
For the massive wave equation, the cylinder spacetimes seem not even to be 
benign:  A single massive particle leaving any spacelike surface $S$ 
can be aimed to return to the surface at a point different from the 
one from which it left. \\
{\em Problem}  Prove (or disprove) the conjecture that the flat
cylinders, given by the identifications (\ref{id}), are not benign
for the massive scalar wave equation.

\begin{figure}
\includegraphics[height=3in]{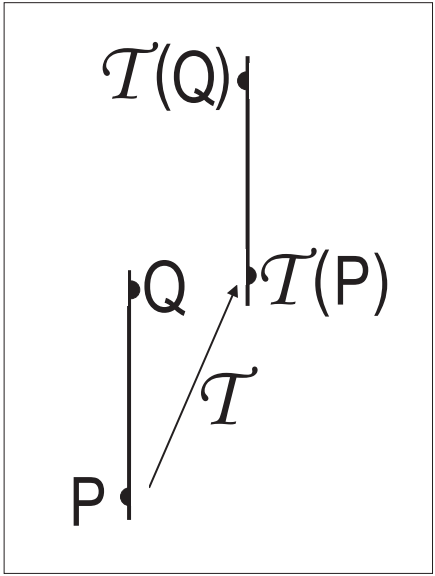}
\caption{\label{slits} A simple spacetime with CTCs and a generalized Cauchy 
surface is shown in this figure.  Two parallel segments of equal 
length are removed from Minkowski space, two disjoint edges are 
joined to the left and right sides of each slit, and edge points 
related by the timelike translation ${\cal T}$ are then 
identified. }
\end{figure}

A spacetime $M,g$ that avoids the problems so far encountered --- the
grandfather paradox and the lack of candidate Cauchy surface --- is akin
to spaces discussed by Geroch and Horowitz \cite{gh} and by
Politzer\cite{politzer}.  Heuristically, as illustrated in
Fig.~\ref{slits}, one removes from Minkowski space two parallel,
timelike slits that are related by translation along a
different timelike direction. The inner edges of the two slits 
are then glued; and the outer edges of the two slits are similarly 
glued. 
The formal construction and details of the initial value problem 
outlined below are given in Friedman and Morris \cite{fm97}.  
Analysis of the initial value problem for a related spacetime with
spacelike slits is given by Goldwirth et al.\cite{gppt} 

Because corresponding points on the left and right slits are related by
a timelike translation, closed timelike curves extend from the left to
the right slit, e.g., from the point labeled $Q$ on the left to the
identified point labeled ${\cal T}(Q)$ on the right.  The {\em
dischronal} region $A$ of a spacetime is the set of points through
which there are closed timelike or null curves. Here it is a bounded
region within the intersection of the past light cone of the top slit
endpoint and the future light cone of the bottom slit endpoint.  A
hypersurface $\Sigma$ that lies in the past of $A$ and is a Cauchy
surface of Minkowski space is an obvious candidate for a generalized
Cauchy surface of $M,g$.  In fact, it is easy to see that initial data
in $L_2(\Sigma)$ leads to a solution in $L_2(M)$.  In the past of $A$,
solutions to the massless wave equation can be written as the sum
$f(t-x) + g(t+x)$ of a right-moving and left-moving solution.  To
obtain a solution in the spacetime $M$, one simply propagates left
moving data that encounters the slit in the obvious way.  For example,
if a left-moving wave enters the left slit at $Q$, it emerges unaltered
from the right slit at ${\cal T}(Q)$.  The solution is unique. But it is
discontinuous along future-directed null rays that extend from the
endpoints of the slits, because the result of the wave propagation is
to piece together solutions from disjoint parts of the initial data
surface.  For example, the right-going solutions on adjacent sides of
right-directed null ray from any endpoint are Minkowski space solutions
obtained from data that came from segments of $\Sigma$ that are not
adjacent.

The existence of solutions in $L_2$, however, is not a generic property of 
two-dimensional spacetimes with a Cauchy horizon.  If, for example,
the two timelike slits were not parallel, the resulting spacetime would 
have an unstable Cauchy horizon: If $\Sigma$ is a Cauchy surface for 
the past of $A$, data on $\Sigma$ leads to a solution that diverges on 
the boundary of the past of $A$. The paradigm for this 
generic case is {\em Misner space} \cite{misner,hawkellis,thorne94}.

Misner space can be constructed by identifying the edges of a strip of
Minkowski space between two parallel {\em null} lines.  As in the 
previous example the CTCs of Misner
space are confined to a spatially bounded region, and one can ask whether
spacelike surfaces lying to the past or future of the dischronal region
$A$ are generalized Cauchy surfaces. \\

To construct the space, let $u = t-x$, $v=t+x$, and consider the
null strip $u_0 < u < B u_0$, where $u_0>0$, and a boost 
${\cal B}$ of Minkowski space corresponding to velocity $V>0$ is given by
\be
u\rightarrow B u,\qquad v\rightarrow B^{-1} v,
\ee
with
\be
B = \sqrt{\frac{1+V}{1-V}}
\ee
Points at the boundaries of the strip are identified after a boost:
\be
(u_0,v)\equiv (B u_0, B^{-1} v)
\ee
Identified points are spacelike separated for $v<0$ (e.g., $P$ and
$P'$ in Fig. 2), null separated at $v=0$ (e.g., $N$ and $N'$), and
timelike separated for $v>0$ (e.g., $Q$ and $Q'$).  Closed timelike
curves (e.g., the segment $QQ'$) thus pass though each point of the
region $v>0$.
\begin{figure}
\includegraphics[height=3in]{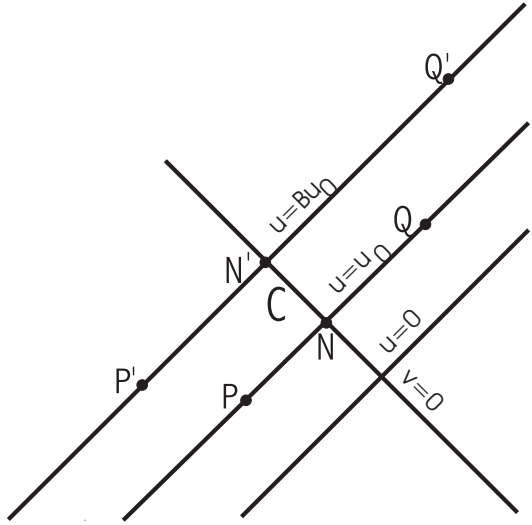}
\caption{\label{fig:misner} Misner space is the region between the two null rays $u=u_0$
and $u=Bu_0$, with points of the null boundaries identified by the boost
${\cal B}$.  The curve $C = NN'$ is a chronology horizon, a closed null
geodesic that separates the dischronal region above it from the
globally hyperbolic spacetime to its past. }
\end{figure} 
Misner space has a single closed null geodesic, $C = NN'$, and the past
${\cal P}$ of $C$ is globally hyperbolic. The future of $C$ is
dischronal, so $C$ is a chronology horizon, a Cauchy horizon
that bounds the dischronal region.  Initial data for the scalar-wave
equation can be posed on a
Cauchy surface $\Sigma$ of ${\cal P}$, but solutions have divergent energy on 
the chronology horizon.

 This globally hyperbolic past part of
Misner space can obtained from a 1-dimensional room whose walls are
moving toward each other -- by identifying left and right walls at the
same proper time read by clocks  on each wall (see, e.g., Thorne
[1994]). This construction makes it obvious that light rays are boosted 
each time they traverse the space, in the same way that a light 
ray is boosted when reflected by a moving mirror.  

The reason solutions diverge is then clear in the geometrical optics limit.
A light ray $\gamma$, starting from $\Sigma$, loops about the space and
is boosted each time it loops.\footnote{That is, trajectories of a (locally-defined)
timelike Killing vector cross the null geodesic at a sequence of
points. The Killing vector can be used to compare the affine parameter
at succesive crossing points by time-translating a segment of the
geodesic to successively later segments.  Compared in this way, the
affine parameter of a given segment will will be less than that of the
next segment by the blueshift factor $[(1+V)/(1-V)]^{1/2}$.}  
Because $\gamma$ loops an infinite number of times before reaching
${\cal C}$, its frequency and energy diverge as it approaches the
horizon.  The ray $\gamma$ is an incomplete geodesic: It reaches the
horizon in finite affine parameter length, because each boost decreases
the affine parameter by the blueshift factor $ [(1+V)/(1-V)]^{1/2}$,
with the velocity of the boost $V$. 

This behavior is not unique to Misner space: A theorem due to Tipler
\cite{tipler} shows that geodesic incompleteness is generic in
spacetimes like Misner space in which CTCs are ``created'' --
spacetimes with a dischronal region to the future of a spacelike
hypersurface.  And a similar argument by Hawking underpins the classical
part of his Chronology Protection Conjecture. \cite{hawk} (See also 
Chrusciel and Isenberg \cite{ci}, who show that the generic, 
compactly generated horizon has generators whose structure is more 
complex than that considered by Hawking.)

When the horizon is not compactly generated, classical fields need not
diverge, and a class of Gott spacetimes \cite{gott} serve as an
example.  Cutler \cite{cutler} shows that a spacelike hypersurface
$\Sigma$ extends to spatial infinity and lies to the past of the
dischronal region.  Here CTCs run to spatial infinity. These
characteristics hold for the particular Gott spacetime introduced here,
but an additional key feature is that its covering space is
three-dimensional Minkowski space (with images of the string
singularities removed).  Carinhas
\cite{carinhas} has shown for the massless scalar wave equation that
data on $\Sigma$ satisfying suitable asymptotic conditions leads to
solutions on a set of Gott spacetimes (see also Boulware \cite{boulware}).

\section{Existence and uniqueness theorems for some four dimensional 
spacetimes with CTCs}

In four spacetime dimensions, existence and uniqueness theorems have
been proved for a class of stationary, asymptotically flat
spacetimes.\cite{fm97,bachelot}  In these spacetimes, the dischronal
region is bounded in space, but there is no Cauchy horizon and CTCs are
always present.  Because the spacetimes are asymptotically flat, one can
define future and past null infinity ${\mathscr I}^\pm$.\cite{hawkellis}
In Minkowski space ${\mathscr I}^-$ is a 
generalized Cauchy surface for massless wave equations, and the goal
here is to show that ${\mathscr I}^-$ is a also generalized Cauchy
surface for a class of spacetimes with CTCs.
\\
 
   We first review in some detail work by Friedman and Morris \cite{fm97}
on spacetimes with topology ${\cal N} = {\cal M}\times {\mathbb R}$, 
where ${\cal M}$ is a hyperplane with a handle (wormhole) attached: 
${\cal M} = {\mathbb R}^3\#(S^2\times S^1)$.   
The metric $g_{\alpha\beta}$ on ${\cal N}$ is smooth
($C^\infty$), and, for simplicity in treating  the asymptotic behavior
of the fields, we will assume that outside a compact region ${\cal R}$
the geometry is flat, with metric $\eta_{\alpha\beta}$.

\begin{figure}
\includegraphics[height=3in]{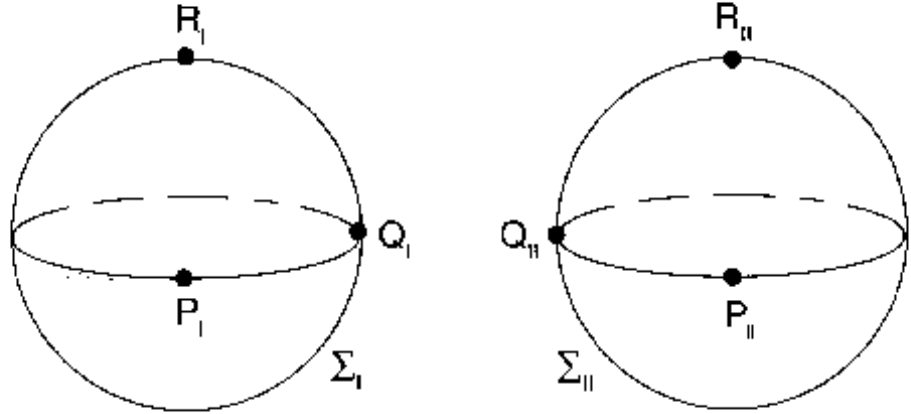}
\caption{\label{tunnel} An orientable 3-manifold $M$ is constructed by identifying
points of $\Sigma_I$ and points of $\Sigma_{II}$ that are labeled by
the same letter, with subscripts I and II. }
\end{figure}

One can construct the 3-manifold ${\cal M}$ from ${\mathbb R}^3$ by
removing two balls and identifying their spherical boundaries,
$\Sigma_I$ and $\Sigma_{II}$, as shown in Fig.~\ref{tunnel}.  The
sphere obtained by the identification will be called the ``throat'' of
the handle. (Its location is arbitrary:  After removing any sphere,
$\Sigma$, from the handle of $\cM$ one is left with a manifold
homeomorphic to ${\mathbb R}^3\backslash(B^3\# B^3)$, whose boundary is
the disjoint union of two spheres.) One can similarly construct the
spacetime $\cal N$ from ${\mathbb R}^4$ by removing two solid cyliders
and identifying their boundaries $C_I$ and $C_{II}$.  We will denote by
$\cT$ the map from $C_I$ to $C_{II}$ that relates identified points.
For the spacetimes we will consider, the identified points will be
timelike separated.\\

A static metric on $\cN$ is given by 
\begin{equation}
 g_{\alpha\beta} = - e^{-2\nu}t_\alpha t_\beta + h_{\alpha\beta}, \label{gab1}
\end{equation}
where $h_{\alpha\beta} t^\beta = 0$.  

If the Minkowski coordinate $t$
is extended to $\cN\backslash C$ by making $\cM_t$ a $t$=constant
surface, then $t^\alpha \nabla_\alpha t = 1$, $\nabla^\alpha t = -
e^{-2\nu}t^\alpha$, and the metric (\ref{gab1}) can be written on
$\cN\backslash C$ in the form
\begin{equation}
g_{\alpha\beta} = - e^{2\nu}\partial_\alpha t \partial_\beta t 
+ h_{\alpha\beta}.\label{gab2}\end{equation}
It will be convenient to single out a representative hypersurface,
\begin{equation}
\cM := \cM_0. \label{cm}
\end{equation} 
We will denote by $h_{ab}$ the corresponding spatial metric on $\cM$;
that is, $h_{ab}$ is the pullback of $h_{\alpha\beta}$ (or
$g_{\alpha\beta}$) to $\cM$.

We consider the wave equation   
\begin{equation}
\sqr87\ \Phi\equiv \nabla^\alpha \nabla_\alpha \Phi = 0, \label{scwaveeq}
\end{equation}
for a massless scalar field $\Phi$.

Initial data in Minkowski space on a future null cone is simply a 
specification of the field $\Phi$ on that cone.  On ${\mathscr I}^-$, 
the field itself vanishes, but the field rescaled by a radial 
coordinate is finite; initial data on ${\mathscr I}^-$ 
can then be written in terms of the standard ingoing null coordinate $v$, 
radial coordinate $r$, and unit radial vector $\bf\hat r$ as 
\be
 f(v,\hat{\bf r}) = \lim_{r\rightarrow\infty} r\Phi(v,r\hat{\bf r}).
\ee  
The same data on ${\mathscr I}^-$ determines a solution to the wave 
equation in the spacetimes ${\cal N}, g$.

{\bf Proposition 1}.\cite{fm97}  {\sl For almost all spacetimes 
${\cal N}, g$ of the kind just described (for almost all parameters $\tau$), 
the following existence theorem holds. Let $f$ be initial data on 
${\mathscr I}^{-}$ for which f and all its derivatives are in 
$L_2({\mathscr I}^{-})$.  Then there exists a solution $\Phi$ to the scalar wave
equation that is smooth and asymptotically regular at null and spatial
infinity and that has $f$ as initial data.}

Because the geometry is static, we can express
solutions as a superposition of functions with harmonic time
dependence. The fact that there is no foliation by spacelike slices 
leads to a lack of orthogonality of the eigenfunctions, and 
the spectral theorem cannot be used.  Instead we explicitly prove 
convergence of a superposition of the form 

\label{harm}
\begin{equation}
\Phi(t,x) = \int d\omega\ \phi(\omega,x)\ e^{-i\omega t}.
\end{equation} 

Here $x$ is naturally a point of the manifold of trajectories of $t^\alpha$,
but we can identify it with a point of a simply connected spacelike
hypersurface $\cM $, with spherical boundaries $\Sigma_I$ and
$\Sigma_{II}$.  Let $(t, x_I)$ and $(t+\tau, x_{II})$ be
points of $\cN\backslash C$ that are identified in $\cN$.  Continuity 
of $\Phi$ and its normal derivative at the identified points is 
expressed by 
\begin{equation}
\Phi(x_{II}) = \Phi(x_I)\label{kgbca}\end{equation}
\begin{equation}
\hat n_{II}\cdot \nabla\Phi(x_{II}) =
-\hat n_I\cdot\nabla\Phi(x_I).
\label{kgbcb}
\end{equation}

The harmonic components of $\Phi$ on $\cN$ can be regarded as fields 
$\phi(\omega,x)$ on $\cM$ satisfying elliptic 
equations of the form 
\begin{equation}
(\omega^2 + \cL)\phi = 0, 
\label{lphi}
\end{equation}
where $\cL$ can be defined by 
the action of $\nabla_\alpha \nabla^\alpha$ on time independent 
fields $f$ on $\cN$:
\begin{equation}
\cL = e^{\nu}D^a e^\nu D_a, \label{defl2}
\end{equation}
and $D_a$ is the covariant derivative of the 3-metric $h_{ab}$ on $\cM$.
The major difficulty lies in the fact that, because the boundary conditions
(\ref{kgbca},\ref{kgbcb}) involve a time-translation by Killing parameter 
$\tau$, the corresponding boundary
conditions on the harmonic components $\phi$ depend on the frequency $\omega$, 
via a phase $\eta = \omega\tau$:
\begin{equation}
\phi(\omega,x_{II}) =  e^{i\eta} \phi(\omega,x_I),
\label{kghba}
\end{equation}
\begin{equation}
\hat n_{II}\cdot \nabla\phi(\omega,x_{II}) =
- e^{i\eta} \hat n_I\cdot\nabla\phi(\omega, x_I).
\label{kghbb}
\end{equation}

As a result, eigenfunctions associated with different frequencies are 
eigenfunctions of different operators; they are not orthogonal, and their 
completeness is not guaranteed by the spectral theorem. 

Instead, the following steps outline the construction of a solution. \\

1. For a fixed value $\eta$ of the phase,  
{\sl the operator $\cL_{\eta}$ with boundary conditions
(\ref{kghba},\ref{kghbb}) is self-adjoint on the space $L_2(\cal M)$
with domain ${\mathbb H}_2$.}\\

Here the boundary conditions enforce the symmetry of the operator
\begin{equation}
\langle f\mid\cL_{\eta} g\rangle = \langle\cL_{\eta} f\mid g \rangle= 
\langle h\mid g
\rangle,  
\label{sa3}
\end{equation}
by requiring that the current entering $\Sigma_I$ coincide with the 
current leaving $\Sigma_{II}$:
\begin{equation}
\int_{\Sigma_{II}} dS_a e^{-\nu}\ (\bar f D^a g - \bar g D^a f)
 + \int_{\Sigma_I} dS_a e^{-\nu}\ (\bar f D^a g - \bar g D^a f).
\label{sa2}
\end{equation}   

2.  Eigenfunctions exist whose incoming part coincides with the 
incoming part of a plane wave for each wavevector $k$.   
These are solutions $F(\eta,k,x)$ to 
Eq.~(\ref{lphi}) that, for $r>R$, have the form
\begin{equation}
F =(2\pi)^{-3/2}e^{ik\cdot x}+\mbox{outgoing waves}.
\label{fk},
\end{equation}
Existence is proved, following Wilcox \cite{wilcox}, by the limiting 
absorption method:  One
adds an imaginary part to the frequency $\omega = |k|$.  
Because ${\cal L}_{\eta}$ is self-adjoint, ${\cal L}_{\eta} +
i\epsilon$ is invertible in $L_{2}$.  One can rewrite the homogeneous 
equation ${\cal L}_{\eta+i\epsilon} F = 0$, for $F$ with asymptotic 
behavior (\ref{fk}), as an inhomogeneous equation 
${\cal L}_{\eta+i\epsilon} F_{\rm out} = \rho$, with 
$F_{\rm out}$ purely outgoing for $r>R$.  The sign of the imaginary part 
of the frequency enforces an outgoing solution, and $F$ is then found 
from the limit, as the imaginary part goes to zero, of a family of 
functions $F_{\rm out}$ in $L_2$.   

3. In flat space, the solution for data $f$ on  ${\mathscr I}^- $ can be written in 
terms of the fourier transform $\widetilde f$ of $f$ in the form 
 \[ \Phi(t,x) = \mbox{Re}\ 2
 		\int d^3 k\ a(k) e^{i(k\cdot x -\omega t)} , \] 
where 
\[
\tilde f (\omega,\hat r) = i\omega a(-\omega \hat r),\quad \omega\geq 0. 
\label{fa1}
\]
Here, $e^{ik\cdot x} $ is replaced by $F(\eta = \omega\tau, k,x) $, 
and one shows convergence 
of their superposition,  
\be 
\Phi(t,x) = \mbox{Re}\ 2 \int dk  F(\eta = \omega\tau, k,x) a(k) .
\label{phiak}
\ee

Although one cannot directly use the spectral theorem, convergence of the 
integral does rely on a related unitarity relation for the eigenfunctions 
$F$ of the Hermitian operator ${\cal L}_\eta$ for fixed 
boundary phase $\eta$.  That is, regarded as a map from
$L_2(\cM)$ to $L_2({\mathbb R}^3)$, $F$ is norm-preserving: 
\be
 \| g \|_{L_2(\cM)} 
 = \left|\left| \int dV_x F(\eta, k,x) g(x) \right|\right|_{L_2({\mathbb R}^3)}
\ee

This allows us to bound the norm of a truncated fourier transform of $F$:  
Let $\chi$ be a smooth step function, satisfying 
\[ \chi(x) = \left\{\begin{array}{ll}	 0, & r>R+\epsilon \\
				1,  & r<R 
		\end{array}\right.\]
\begin{equation}
\hat F(\eta,k,y):= \int dV_x e^{-\nu}\ F(\eta,k,x)\chi(x) 
\label{hatF}\end{equation}
has uniformly bounded norm in $k$-space,
\begin{equation}
\left|\left|\hat F(\eta,\cdot,y)\right|\right|_{L_2({\mathbb R}^3)}\leq CR^{3/2},
     \ \forall \eta,y.
\label{normf1}
\end{equation}  
From this uniform bound, one can show
\begin{equation}
\int d\tau dk dy\ 
{\omega^{2n}|\hat F(\omega\tau,k,y)|^2
	\over(1+\omega^2)^n (1+y^2)^{3/2+\epsilon}} < \infty.	\nonumber
\end{equation}
This inequality, in turn implies 
\begin{eqnarray}
& \omega^n\hat F(\omega\tau,k,y)
\in L_2(I)\otimes L_{2,-n}({\mathbb R}^3)\otimes L_{-3/2-\epsilon}({\mathbb R}^3)\nonumber\\
\Longrightarrow\ &\nabla^n F(\omega\tau,k,x)
\in L_2(I)\otimes L_{2,-n}({\mathbb R}^3)\otimes H_{-3/2-\epsilon}(\cM_D)\nonumber\\
\Longrightarrow\ & F(\omega\tau,k,x)
\in L_2(I)\otimes L_{2,-n}({\mathbb R}^3)\otimes H_{n-3/2-\epsilon}(\cM_D).\nonumber\\
\label{ineqa}
\end{eqnarray}
Thus, for almost all $\tau$, 
\begin{eqnarray}
& F(\omega\tau,k,x)\in L_{2,-n}({\mathbb R}^3)\otimes H_{n-3/2-\epsilon}(\cM_D)
\nonumber\\
\Longrightarrow\ &\int dk\ a(k) F(\omega\tau,k,x)\in H_{n-3/2-\epsilon}(\cM_D)
\label{ineqb}
\end{eqnarray}
Finally, $f\in H_n({\mathscr I}^-)$, all $n$, implies $\widehat f \in L_{2,-n}$, 
all $n$, whence $\Phi$ given by Eq. (\ref{phiak}) is smooth.  

4.  Asymptotic regularity of $\Phi$ follows from its explicit 
form for $r>R$ in terms of the value of $\Phi$ and $\nabla \Phi$ at $r=R$.
That is, one can use the flat-space Green function to write $\Phi$ outside
$r=R$.   

    More recently, Bachelot \cite{bachelot} has proved a similar existence
theorem and a strong uniqueness theorem for another family of
stationary, four-dimensional spacetimes that are flat outside a
spatially compact region.  These spacetimes have Euclidean topology and
their dischronal regions have topology (solid torus) $\times {\mathbb R}$.  The
metric is axisymmetric, with one free function $a$ that describes the
tipping of the light cones in the direction of the rotational Killing
vector $\partial \phi$.

\be
g = - (dt-a\ d\phi)^2 + dr^2 + r^2 d\theta^2+ r^2\sin^2\theta^2 d\phi^2.
\ee
Circles about the axis of symmetry are CTCs when $\partial_\phi$ is
spacelike --- that is, when $r\sin\theta <a$.  By choosing $a=0$
outside a torus, one can restrict CTCs to the interior of a smaller
torus.  Again data at ${\mathscr I}^-$ for $\Phi$ yields a smooth,
asymptotically regular solution $\Phi$, and Bachelot shows that $\Phi$
is unique.  This is a significantly stronger result than the weak 
uniqueness obtained for the wormhole spacetimes described above, 
and it suggests that a strong uniqueness theorem should hold for
those spacetimes as well.   

\section{Conjectures for more general four-dimensional spacetimes}

As noted in Sect. {\ref{2d}} four-dimensional spacetimes that have
Cauchy horizons and satisfy the null energy condition are geodesically
incomplete.  In two dimensions, an incomplete null geodesic that
appproaches a closed null geodesic as it approaches the chronology
horizon leads to instability of that horizon.  In four dimensions,
however, an incomplete null geodesic $\gamma$ does not always imply
that the chronology horizon is unstable.  This is because there may be
only a set of measure zero of such geodesics, so that the energy may
remain finite on the chronology horizon.  For the time-tunnel
spacetimes considered in refs \cite{mt,mty,fmetal}, a congruence of null
rays initially parallel to $\gamma$ spreads as the rays approach the
chronology horizon. When the spreading of the rays overcomes the
successive boosts (when the fractional decrease in flux is greater than
the fractional increase in squared frequency), the horizon is stable in
the geometrical optics approximation, and we will call it {\em
optically stable}.  (A precise, but long-winded definition of optical
stability is given in Ref. \cite{fm97};

a similar definition, applicable in a more restricted context, is given by Hawking\cite{hawk}).
Because the instability of the chronology horizon (or of the spacetime to its future) appears to be
the obstacle to existence of solutions for data on candidate generalized Cauchy surfaces, 
we are led to a conjecture that relates optical stability to the existence of solutions.  

\noindent
{\sl Existence Conjecture}.  Let $\cN, g$ be  a smooth, asymptotically
flat spacetime for which past and future regions ${\cal P} = \cN \backslash
J^+ ( A)$ and ${\cal F} = \cN \backslash J^- (A) $ of a compact
4-dimensional submanifold $A$ are globally hyperbolic. If $\cN, g$ is
optically stable, solutions to massless wave equations (for
scalar, Maxwell, and Weyl fields) exist on $\cN, g$ for smooth data 
on a Cauchy surface for ${\cal P}$.  

A conjecture relating uniqueness for massless fields to
uniqueness in a geometic-optics sense is easier to formulate.

\noindent
{\sl Uniqueness Conjecture}.  Again let $\cN, g$ be  a smooth,
asymptotically flat spacetime for which past and future regions 
${\cal P} = \cN \backslash J^+ ( A)$ and ${\cal F} = \cN \backslash J^- (A)$
of a compact 4-dimensional submanifold $A$ are globally hyperbolic.
Let $S_\pm$ be Cauchy surfaces for $\cN\backslash J^\pm(A)$.  If all but a set
of measure zero of null geodesics intersect $S_+$ and $S_-$, then
solutions to massless wave equations on $\cN$ are unique for initial
data on $S_-$ (and for initial data on $S_+$).

If one omits the restriction on null geodesics, uniqueness fails: It is
not difficult to construct spacetimes satisfying the remaining
conditions of the conjecture for which solutions to the massless scalar
wave equation have support on a compact region.\cite{fm97}  One example
begins with a 4-torus with flat Lorentz metric chosen to make two of
the generators null and the other two spacelike. The metric allows a
nonzero plane-wave solution whose support is not the entire torus.  On
can smoothly glue the torus to an asymptotically flat Lorentian
spacetime without altering the metric on the support of the scalar
field.  \\

For no asymptotically flat spacetime in 4-dimensions, in which CTCs are
confined to a compact region, am I aware of a rigorous demonstration
that finite-energy solutions to the scalar wave equation do exist for
arbitrary initial data, or that solutions are unique. 

Still less is known about interacting fields.  

The well-known billiard-ball examples of Echeverria 
{\sl et al.}\cite{fmetal,ekt91} are the basis for our present intuitive 
understanding.  These examples exhibit a multiplicity of solutions for
the same initial data, suggesting that uniqueness in spacetimes with CTCs is likely to
hold only for free or weakly interacting fields. Because solutions
seem always to exist for the billiard ball examples in the spacetimes
they considered, it may be that classical interacting fields  
have solutions on spacetimes for which solutions to the
free field equations exist.  

Fewster, Higuchi and Wells \cite{fhw} looked at a model of an interacting 
field theory in which space is discrete, and time is identified to 
obtain a discrete version of 2-dimensional Minkowski space with two horizontal 
slits removed and opposite edges of the slits identified. The field $\psi$
satisfies an equation of the form 
\[ \partial_t \psi = L\psi + \lambda \psi^\dagger \psi\psi ,\]
where $L$ is a linear operator, and $\lambda$ is real.  

Fewster et al. find that solutions exist for arbitrary data and arbitrary 
$\lambda$ and that they are unique for small $\lambda$.  For large 
$\lambda$, however, uniqueness is lost. \\

An obvious question is whether generalized Cauchy surfaces for free 
fields similarly serve as generalized Cauchy surfaces for weakly interacting
fields; and whether, as the toy models suggest, uniqueness is fails 
above some critical value of the interaction parameter.

\end{document}